\documentclass[11pt]{article}
\usepackage[margin=1in]{geometry}
\usepackage{cite}

\usepackage{blindtext}
\usepackage{amsmath,amssymb,amsfonts}
\usepackage{algorithmic}
\usepackage[algo2e,linesnumbered,vlined,algoruled]{algorithm2e}
\usepackage{graphicx}
\graphicspath{{figures/}}
\usepackage{textcomp}
\usepackage{url}
\usepackage{amsthm}
\usepackage{array}
\usepackage{enumitem}
\usepackage[frozencache=true,cachedir=.]{minted}

\setlist{leftmargin=4mm}
\usepackage{xcolor}

\usepackage[breaklinks=true, colorlinks=true, linkcolor={blue!90!black}, citecolor={blue!90!black}, urlcolor={blue!90!black}]{hyperref}

\usepackage[normalem]{ulem} 
\newcommand\hcancel[2][black]{\setbox0=\hbox{$#2$}\rlap{\raisebox{.45\ht0}{\textcolor{#1} {\rule{\wd0}{1pt}}}}#2} 
\newcommand{\replace}[2]{{\color{red}\color{black}{\color{black}#2\color{black}}}}
\newcommand{\replacemath}[2]{{\hcancel[red]{}{}{\color{black}#2\color{black}}}} 

\usepackage{multicol}
\usepackage{stfloats}

\usepackage{comment}
\usepackage{xspace}
\newcommand{\libE}{libEnsemble\xspace}

\usepackage{authblk}
\begin{document}

%
\title{libEnsemble: A Library to Coordinate the Concurrent \\Evaluation of Dynamic Ensembles of Calculations}

\author[]{Stephen Hudson}
\author[]{Jeffrey Larson}
\author[]{John-Luke Navarro}
\author[]{Stefan M.~Wild}

\affil[]{Argonne National Laboratory\\ Mathematics and Computer Science Division\\

{\rm \texttt{\{shudson,jmlarson,jnavarro,wild\}@anl.gov}}
}
\date{}

\markboth{}%
{Hudson \MakeLowercase{\textit{et al.}}: libEnsemble}
\maketitle

\begin{abstract}
Almost all applications stop scaling at some point; those that don't 
are seldom
performant when considering time to solution on anything but
aspirational/unicorn resources. Recognizing these tradeoffs as well as greater
user functionality in a near-term exascale computing era, we present libEnsemble, a
library aimed at particular scalability- and capability-stretching uses. libEnsemble
enables running concurrent instances of an application in dynamically allocated
ensembles through an extensible Python library.
We highlight the structure, execution, and capabilities of the library on leading
pre-exascale environments as well as advanced capabilities for 
exascale environments and beyond.

\end{abstract}

\section{Introduction and \libE Aims}\label{sec:introduction}

Scientific calculations are able to use increasing amounts of computational
resources. Even so, the calculation's time to solution will eventually stop
improving as additional resources are used. 
Many applications would still want to efficiently use  resources such as emerging exascale computers; running concurrent instances of the 
calculation is a natural next step for goal-oriented application problems.

We refer to such instances of calculations for a shared goal as members of an \emph{ensemble}.  When considering computations from an ensemble-based view, many tradeoffs arise. For example, some algorithmic approaches may scale to far greater concurrency than others. A case of the former is brute-force computations (e.g., numerical integration by random sampling or by evaluating a deterministic low-discrepancy sequence): given an ensemble size corresponding to the concurrency level, the members of the ensemble can be formed a priori into a predetermined set and then be given out for evaluation. On the other hand, for many performance measures other than utilization or in less-than-ideal settings (e.g., when the calculations show nontrivial variability in their individual runtimes, when calculations may fail in a nonuniform way), such a brute-force ensemble approach is clearly unsatisfactory. That is, even if the resulting computation scales (near-)perfectly with ensemble size, the baseline (ensemble size of one) case is so poor as to render such naturally scalable approaches inferior for imagined concurrency levels. At the other extreme are entirely sequential approaches, which assume that a calculation is saturating available computational resources and that subsequent calculations may be optimally determined based on the result of all prior calculations.

In this paper we introduce \replace{}{\libE \cite{libensemble-man}}, a Python library that seeks to empower users to address such tradeoffs. 
\replace{}{\libE originated in the confluence of 
\begin{itemize}
    \item designing outer-loop optimization, sensitivity analysis, and uncertainty quantification solvers around PETSc~\cite{petsc-user-ref,mills2020performanceportable} capabilities; 
    \item exploring 
    high-performance Python functionalities such as petsc4py\footnote{See \url{https://pypi.org/project/petsc4py/}, which has 
    subsequently been incorporated in PETSc via the \texttt{--with-petsc4py} option}
    and mpi4py~\cite{Dalcin2008}; and
    \item developing scalable implementations of APOSMM.
\end{itemize}
The latter represents an asynchronous multistart optimization 
method~\cite{LW16,LarWild14} that requests points either randomly generated or arising in various local
optimization runs. In most use cases, these points parameterize computational simulations 
run on HPC systems at Argonne.} 

\replace{}{Additional design motivations included consideration  of cases where single instruction, multiple data (SIMD) workflows may hamper meaningful progress toward a stated goal. As an example, 
deep/supervised machine-learning methods often presume that a fixed batch size of
training data is used in every iteration. Such methods are rooted in
resource-constrained settings where model evaluations naturally exploit SIMD benefits. 
Constraining oneself to such a
workflow may result in more iterations, instructions, or data accesses than
approaches where a batch size is determined adaptively~\cite{BollapragadaICML18,Li2020}.}

The design aims of \libE include the following:
\begin{enumerate}
    \item Tightly coordinated communication between a small number of highly configurable functions 
    \item Ability for resource-\replace{dependent}{aware} ensemble \replace{configuration}{execution} 
    \item Significant control over output and future ensemble members
    \item Extreme portability, targeting automated scaling from laptops to \replace{supercomputers}{exascale systems}
\end{enumerate}


Below we expand on these points and outline this paper. 


In order to tightly coordinate parallel ensembles of 
calculations across a variety of scales and communication media,
\libE employs a basic \textit{manager-workers} paradigm. As detailed in \S\ref{subsec:simgenchar}, workers call Python 
\textit{generator} and \textit{simulator} functions (or wrappers)
to perform any 
type of computation and exchange data with the manager to determine and initiate future ensemble 
members. Such a paradigm enables the support of multiple communication protocols (\S\ref{subsec:runninglibeparallel}). Generator, simulator, and \emph{allocation} functions together are \textit{user functions}
(\S\ref{sec:user_funcs}) and form the basis of all computations performed with \libE.

\replace{The third building block of \libE is a means for the }{To address the second aim, \libE uses a} manager to allocate work to the workers via a  resource-aware \textit{allocation} function
(\S\ref{subsec:allocf}) designed to enable ensemble evaluations that can adapt to available resources. 
\replace{\libE is in part motivated by cases where single instruction, multiple data (SIMD) workflows may hamper meaningful progress toward a stated goal. As an example, 
deep/supervised learning methods often presume that a fixed batch size of
training data is used in every iteration. Such methods are rooted in
resource-constrained settings where model evaluations naturally exploit SIMD benefits. 
Constraining oneself to such a
workflow may result in more iterations, instructions, or data accesses than
approaches where a batch size is determined adaptively~\cite{BollapragadaICML18,Li2020}.}{} 
The allocation of \libE work to workers via an allocation function is facilitated by a \textit{history} array (\S\ref{subsec:history}) of 
all pending and completed computations performed by workers via generator and simulator 
functions. This allocation can readily account for the availability of resources relative to history-based insights.

\libE addresses \replace{settings}{scenarios} where ensemble members \replace{are}{need to be dynamically defined} and
controlled. Classical ensemble-based computations include linear system solve\replace{}{r}s
involving multiple right-hand sides. When a typical direct method is used for
such multiple right-hand-side problems, a single (e.g., factorization-based)
expense is paid in order to benefit the concurrent solution for \emph{all} possible
right-hand sides. Such a direct solution method is a powerful technique that can be
deployed for many complex problems and modern computational science problems
(see, e.g.,~\replace{}{\cite{Amestoy2019,Gunzburger2019}}).
When the expense for effectively precomputing all possible ensemble solutions
is implausible or unnecessary, obvious alternatives include iterative
solvers, such as those deployed in PETSc~\cite{petsc-user-ref} and
Trilinos~\cite{trilinos-website}. \libE is especially designed with
consideration for settings where there may be benefits from ensemble members
using shared resources and data structures (e.g., discretizations shared among
most members) without an assumption that each ensemble's calculation will share
a common control flow. By not restricting ensemble members to rely on
precomputed work, \libE seeks to enable use cases where it is often better to
focus on calculations in dynamically evolving configurations. In
\S\ref{sec:uses} we overview several such use cases, ranging from parameter
space exploration with simulation failure indicators to design optimization to
statistical calibration. In each case, we assume that, by itself, an ensemble
member may not be able to fully benefit from its available resources and hence it
is important to concurrently execute ensemble members.

One obstacle to employing libraries addressing extreme-scale concurrency
afforded by supercomputers is the inability to develop and deploy such
capabilities at intermediate  levels, including a user's laptop.  \libE was
developed as part of the software ecosystem \cite{ECPST2020} for the U.S.
Department of Energy exascale computing project (ECP) and with a
Python-based front end for use on many systems. 
\libE readily allows 
user functions to launch and 
interface with applications on high-performance hardware
(\S\ref{sec:runningonhpc}). \libE's executor can
dynamically split allocated resources (including GPUs) among workers to the user's
specifications and interface with emerging workflow schedulers such as
Balsam~\cite{balsam18}. On the other hand, \libE is available as a PyPI
(\texttt{pip}), Conda, and Spack package. In all such forms, \libE is actively
developed to be interoperable with other exascale-focused
software technologies and is the first Python package incorporated in xSDK, the
Extreme-scale Scientific Software Development Kit~\cite{bartlett2017xsdk}.

\replace{}{\libE is one of a number of extreme-scale 
workflow software packages (e.g.,  RADICAL-Ensemble Toolkit~\cite{radical} and other packages in the ExaWorks project~\cite{exaworks}). \libE primarily distinguishes itself via its generator-simulator paradigm that sidesteps requiring users to define task dependencies in favor of data dependencies between configurable Python user functions; see~\S\ref{sec:user_funcs}. 
This allows the user to focus their attention on function logic. This composable design also lends itself to exploiting the large library of example user functions that are provided with libEnsemble, maximizing code re-use. For example, users can easily select an existing generator function while modifying a simulator function for their specific use case. While libEnsemble provides a complete ensemble toolkit including a task executor interface, its modular design also allows users to plug in components from other workflow packages such as Parsl~\cite{parsl}, resource management libraries like Flux~\cite{Flux}, or various pilot systems.
}

\libE is a growing software package; we highlight current and future development ideas in \S\ref{sec:discussion}.

\begin{figure}[!t]
\centering
\includegraphics{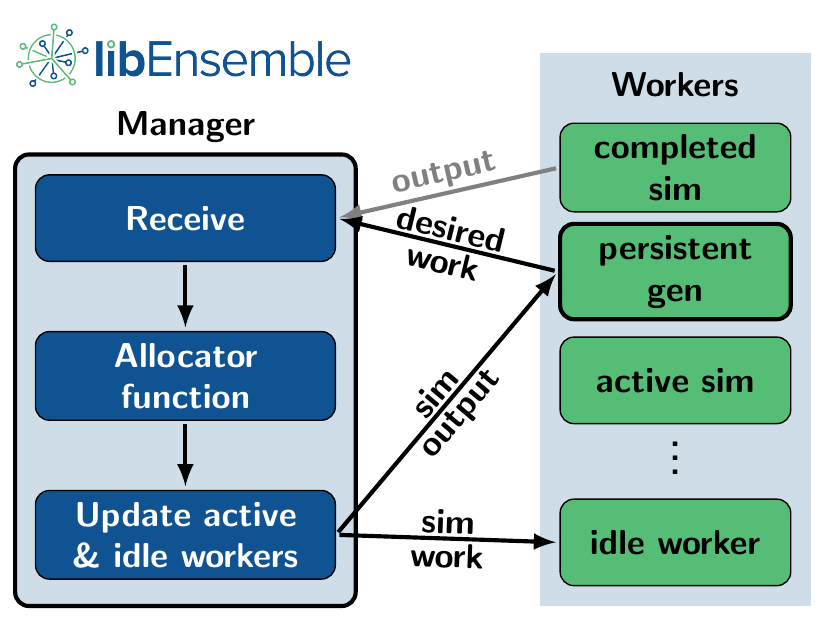}
\caption{Example of data movement between \libE's manager and workers. Here, 
the manager receives output from a completed simulation and also some other work
requested by a persistent generator. This is given to the allocation function, along with 
information about what workers are active or idle. The allocation function determines what 
work should be done and with what resources.
\label{fig:managerworker}}
\end{figure}






\section{\libE User Function Overview}\label{sec:user_funcs}
An early part of \libE's design was the decision to divide ensemble steps into
generator and simulator routines as an intuitive way to express problems and their inherent
dependencies.

\libE was consequently developed to coordinate ensemble computations defined by 
\begin{itemize}
    \item a \textit{generator function} 
    that produces simulation inputs,
    \item a \textit{simulator function} 
    that performs and monitors simulations, and 
    \item an \textit{allocation function} 
    that determines when (and with what resources) the other two functions should be invoked.
\end{itemize}
Since each of these functions is supplied or selected by \libE's users, they are
typically referred to as user functions. User functions need not be written only in Python: they can (and often do) depend on routines from other
languages. The only restriction for user functions is that their inputs and outputs  conform
to the user function API. Therefore, the level of computation and complexity of any user function
can vary dramatically based on the user's needs. 


An example of the interplay between these user functions is given in
Fig.~\ref{fig:managerworker}. In most \libE workflows, users develop and
combine a set of their own user functions or 
develop a single user function to fulfill a particular algorithmic purpose or support a 
specific application. In all cases, users are invited to browse examples distributed
with \libE and modify them for their own needs.

\subsection{Simulator and Generator Function Characteristics}\label{subsec:simgenchar}




On being sent a unit of work by the manager, a worker calls the requested simulator or 
generator function with the following parameters:
\begin{itemize}
    \item \texttt{H}: an assigned selection of the history array (see \S\ref{subsec:history})
    \item \texttt{persis\_info}: a dictionary of per-worker persistent information
    \item \texttt{sim\_specs} or \texttt{gen\_specs}: a dictionary of configuration fields and data types
    \item \texttt{libE\_info}: a dictionary of \libE-specific information\replace{.}{, typically used in persistent functions (see \S\ref{subsec:persistentfunctions})}
\end{itemize}
Inside each function, input data is extracted from the history array selection, and an output 
array is defined based on the user-defined output data types for the function \replace{.}{found in
\texttt{sim\_specs} or \texttt{gen\_specs}.} This array 
is populated by the simulator or generator routines. In the simple simulation function in Fig.~\ref{fig:simfexample}, the output array is populated by matrix norms of input values.

Any level and complexity of computation can be performed. Many functions use 
\libE's executor (\S\ref{subsec:supportuserapps}) to launch applications and monitor results.
Other functions can be written to initiate parallel routines via Python's  \texttt{multiprocessing} module or even detect whether already-scheduled simulations become unnecessary and request 
their cancellation.

Persistent information or structures for use across user function calls can be stored in \texttt{persis\_info}.
\replace{Figure~\ref{fig:genfexample}}{For example, Fig.~\ref{fig:genfexample} depicts} 
\replace{an example}{a simple} generator function that uses 
a NumPy random number stream that is stored in \texttt{persis\_info}
in order to generate random points. This stream can be used in future calls to the generator
function if more controlled random number generation is desired.

Following computations, each simulator or generator function returns three items:
\begin{itemize}
    \item The output NumPy array, populated with evaluations
    \item \texttt{persis\_info}, to be passed back on the next function call
    \item An optional \texttt{calc\_status} integer, typically for logging the status of application runs
\end{itemize}

\begin{figure}[!t]
\centering
\begin{minted}[
frame=lines,
fontsize=\scriptsize,
]
{python}
import numpy as np

def sim_example(H, persis_info, sim_specs, libE_info):

    x = H['x']
    H_o = np.zeros(1, dtype=sim_specs['out'])
    H_o['f'] = np.linalg.norm(x)

    return H_o, persis_info
\end{minted}
\caption{Simple simulator function. The output array is defined to match the 
expected \texttt{\textquotesingle out\textquotesingle} specifications and then populated by calculations.
This array is returned to the manager.}\label{fig:simfexample}
\end{figure}
    
\begin{figure}[!b]
\centering
\begin{minted}[
frame=lines,
fontsize=\scriptsize,
]
{python}
import numpy as np

def random_sample(H, persis_info, gen_specs, _):

    rand_stream = persis_info['rand_stream']
    ub = gen_specs['user']['ub']
    lb = gen_specs['user']['lb']

    n = len(lb)
    b = gen_specs['user']['gen_batch_size']

    H_o = np.zeros(b, dtype=gen_specs['out'])
    H_o['x'] = rand_stream.uniform(lb, ub, (b, n))

    return H_o, persis_info
\end{minted}
\caption{Example generator function. The random stream in
\texttt{persis\_info} is used to generate values and is returned to the manager 
for use in future calls.}\label{fig:genfexample}
\end{figure}
    
\subsubsection{Persistent Functions}\label{subsec:persistentfunctions}

For many more sophisticated generator or simulator functions, the paradigm of initializing, computing,
and returning potentially hundreds of times over the course of a \libE workflow can be inefficient. 
An alternative approach involves structuring the function to receive work, perform computations, and 
return results in a loop, communicating directly with the manager instead of returning results to the 
calling worker. A simulator or generator function structured in this way is referred to as \textit{persistent}, 
with the calling worker considered a persistent worker. Persistent functions take advantage of
the communicator object included in \texttt{libE\_info} and several helper functions to send and receive 
data from the manager.

An example persistent user function depicting the process of looping and directly communicating with the manager is shown 
 in Fig.~\ref{fig:persisgenexample}. Note that the same random uniform sampling is performed here as in 
Fig.~\ref{fig:genfexample}, but the results are immediately sent to the manager via the \texttt{send\_recv()} 
helper function. The manager communicates back a tag and additional work. Persistent functions compare tags from the manager
to know when to exit.

The APOSMM parallel optimization generator function (discussed further in \S\ref{subsec:aposmm}) included with \libE is persistent
so it can maintain and advance local optimization subprocesses based on the results of complete simulations.
Previous nonpersistent versions of APOSMM had to reconstruct the sequence of local optimization runs on every call to the generator.

\begin{figure}[!t]
\centering
\begin{minted}[
frame=lines,
fontsize=\scriptsize,
]
{python}
import numpy as np

from libensemble.tags import STOP_TAG, PERSIS_STOP,
                        PERSIS_FINISHED
from libensemble import send_recv

def persistent_uniform(H, persis_info, gen_specs, 
                       libE_info):
    ub = gen_specs['user']['ub']
    lb = gen_specs['user']['lb']
    n = len(lb)
    b = gen_specs['user']['gen_batch_size']
    rand_stream = persis_info['rand_stream']
    comm = libE_info['comm']

    # Send batches until manager sends stop tag
    tag = None
    while tag not in [STOP_TAG, PERSIS_STOP]:
        H_o = np.zeros(b, dtype=gen_specs['out'])
        H_o['x'] = rand_stream.uniform(lb, ub, (b, n))
        tag, Work, calc_in = send_recv(comm, H_o)
        if calc_in is not None:
            b = len(calc_in)

    return H_o, persis_info, PERSIS_FINISHED
\end{minted}
\caption{Example persistent generator function. The function generates values while communicating directly with the manager every loop.}\label{fig:persisgenexample}
\end{figure}
    
\subsection{Allocation Function Characteristics}\label{subsec:allocf}

As opposed to the simulator and generator functions that are called by the workers, each allocation  function is called by the \textit{manager} as part of the main loop. The call includes
 a list of workers and their statuses, the current state of the history array, and the specifications
available to both generator and simulator functions.

Allocation functions query the status of entries from the history array and
map these to available workers, using a handful of helper functions to select
and define units of work for chosen workers. This allocation is captured in a  \texttt{Work} dictionary.
The allocation function returns the resulting Work dictionary and \texttt{persis\_info} 
to the manager. The manager will iterate over the Work dictionary and send entries to assigned workers.

For allocation functions, as with all user functions, the level of complexity can vary widely.
Various scheduling and work distribution features are available in the existing allocation functions, 
including prioritization of simulations, returning evaluation outputs to the generator immediately or in batch, 
assigning varying resources sets to evaluations, and other methods of fine-tuned control over the data available
to other user functions. \libE also includes an allocation function that allows users to simply distribute 
pre-existing simulator work without needing a generator.


For persistent generator or simulator functions, the allocation function is responsible for initiating 
a given function as persistent and flagging that the associated worker is in ``persistent mode."






\section{Configuring and Running \libE}

Users have a large degree of control over the \libE run and its components via a Python \textit{calling script}, where \libE's parallelism method, the history array, 
user-application registration, and other characteristics are defined.

\subsection{Calling Script}\label{subsec:callingscript}

Any given \libE instance is configured and run from a Python script referred to 
as the calling script. The calling script defines parameters
for the ensemble and specifies the user functions, followed by a call to
the primary \texttt{libE()} initialization function, with each configuration dictionary passed 
in as parameters, as in Fig.~\ref{fig:callingscript}. A \libE calling script can be run by using
an MPI runner (which requires \texttt{mpi4py} communications,\; see \S\ref{subsubsec:mpi})
or serially with either of the other two communication schemes described in \S\ref{subsec:runninglibeparallel}.

\libE does not require a specialized runtime environment or pilot program,
so long as workers are able to issue applications directly (e.g., via an MPI runner).
This is the case on standalone systems and many clusters or when \libE is run on the
dedicated launch nodes of many supercomputers. This keeps the user's installation
and run requirements as simple as possible. In such circumstances, \libE is initiated
simply by running the calling script with Python:

\vspace{5pt}

    \texttt{\$ python libE\_routine.py}.

\vspace{5pt}
Because \libE uses MPI by default, this statement produces only a single MPI
process, leaving none for workers. Specifying enough processes is covered in the next
section.

In scenarios where the direct launching of applications from workers is not possible
or where more advanced scheduling schemes are desired, such as using
multiple resource pools (on the same or different systems), \libE can
employ the 
Balsam (see \S\ref{sec:balsam})
executor.

\begin{figure}[!t]
\centering
\begin{minted}[
frame=lines,
fontsize=\scriptsize,
]
{python}
import numpy as np
from libensemble.libE import libE
from libensemble.sim_funcs import one_d_example
from libensemble.gen_funcs import uniform_random_sample
from libensemble.tools import parse_args

# Parse primary parameters from command-line
nworkers, is_manager, libE_specs, _ = parse_args()

# Specify misc libE settings
libE_specs['save_every_k_sims'] = 10

# Configure simulator and generator
sim_specs = {'sim_f': one_d_example, 
             'in': ['x'], 
             'out': [('f', float)]}
gen_specs = {'gen_f': uniform_random_sample,
             'out': [('x', float, (1,))],
             'user': {'gen_batch_size': 500,
                      'lb': np.array([-3]),
                      'ub': np.array([3])}}

# Specify exit criteria
exit_criteria = {'sim_max': 500}

# Perform the run
H, persis_info, flag = libE(sim_specs, gen_specs, 
                            exit_criteria, libE_specs=libE_specs)
\end{minted}
\caption{Example calling script from \libE's regression tests. \libE and the user functions are
configured via dictionaries, each of which is passed to \texttt{libE()}~to initialize \libE. A complete
list of options for these dictionaries is available on \libE's online documentation~\cite{libensemble-man}.\label{fig:callingscript}
}
\end{figure}

\subsection{Running \libE}
\label{subsec:runninglibeparallel}

\libE supports three main methods for manager--worker communications upon initiation:
\begin{description}
    \item[\S\ref{subsubsec:mpi}] A message passing interface (MPI) via \texttt{mpi4py}~\cite{Dalcin2008} with an MPI implementation such as MPICH2~\cite{Jenkins2014}
    \item [\S\ref{subsubsec:mp}] Python's \texttt{multiprocessing} module for \textit{local} communications
    \item[\S\ref{sec:tcp}] Distributed/cloud-based environments via the transmission control protocol (TCP)
\end{description}
Switching between these methods does not require  workflow code changes, although choosing to use
either MPI or local communications typically implies a preference for what nodes \libE launches on 
and how processes are distributed. By default, \libE launches with MPI.

\subsubsection{MPI}
\label{subsubsec:mpi}

Running \libE with MPI-based communication potentially scales best when running with many workers
on multinode systems, and it allows users to be flexible with how processes
are distributed across nodes (see \S\ref{sec:runningonhpc}). \libE with MPI
is typically initiated via a single \texttt{mpirun} or equivalent statement with
\texttt{python},  providing the number of workers to \texttt{mpirun} and the \libE
calling script as the primary argument to \texttt{python}. MPI rank 0 becomes the 
manager process, and the remaining become worker processes, for example:

\vspace{5pt}

    \noindent\texttt{\$ mpirun -n 64 python libE\_calling\_script.py}.

\vspace{5pt}
Using the MPI executor module\replace{, for example}{} to invoke \replace{a}{an MPI-based} simulations \replace{using parallel resources itself} (see \S\ref{subsec:supportuserapps}) while running \libE with MPI
produces nested MPI processes. \replace{This feature}{Nesting MPI} is not supported by Open MPI but is supported by
MPICH and its derivatives. Users who prefer Open MPI can overcome
this circumstance by using a proxy application launcher such as Balsam.
\libE with MPI requires the installation of \texttt{mpi4py}~\cite{Dalcin2008}.

\subsubsection{Multiprocessing}
\label{subsubsec:mp}
The multiprocessing option uses Python's built-in \texttt{multiprocessing} module for
manager/worker communications and is referred to as \textit{local}
mode. All of \libE's manager and worker processes are initiated on a single
node (\libE must be launched in \textit{central} mode); \libE's executor is needed to use the remaining nodes in
an allocation. On three-tier systems such as ALCF's Theta and OLCF's Summit, \libE
should be run on the launch nodes in local mode, allowing the entire compute
node allocation to be available for application submissions via the executor.

An optional helper function for calling scripts allows users
to run \libE with multiprocessing from the command line:
\vspace{5pt}

    \noindent \texttt{\$ python libE\_routine.py --comms local --nworkers 64}.

\vspace{5pt}

\subsubsection{TCP}
\label{sec:tcp}
Running \libE with TCP communication allows the manager process to run on one system and 
workers to remote systems or nodes. The architecture for this approach is 
based on the Plumbing for Optimization with Asynchronous Parallelism (POAP) 
event-driven framework~\cite{2019-pysot}.


\subsection{History Array}\label{subsec:history}

\libE's manager maintains a NumPy structured array of all input and output
values from each generator and simulator function call during a run. 
User functions must return their computations in NumPy arrays so that they can be slotted
into this history array. The history array maintains \texttt{sim\_ids}
for each unit of work produced by a generator and worker IDs that record which worker was assigned to call 
an associated user function. The Boolean fields \texttt{given} and \texttt{returned} record whether a row in the history has been distributed for simulation and whether results have returned 
from a simulation, respectively.

The current state of the history array is returned to the user on concluding a 
successful \libE run, or it is saved to file in the event of a fatal error or timeout. The
maintenance of these  fields and all computations allows users to restart \libE
routines in the event of a failure by providing a history array.




\subsection{\replacemath{}{Logging, Shutdown, and Error Handling}}

\libE logs ensemble routine information to two files by default. The \texttt{ensemble.log} file contains
overarching run information such as available nodes, where managers and workers are 
initiated, and the status of applications launched via the executor. \libE's logger uses the
standard Python logging levels of \texttt{DEBUG}, \texttt{INFO}, \texttt{WARNING}, and \texttt{ERROR},
with \texttt{INFO} being the default level. \libE also uses an additional custom \texttt{MANAGER\_WARNING}
level between \texttt{WARNING} and \texttt{ERROR}. The logging level can be adjusted in the calling script.

In \texttt{libE\_stats.txt}, a line is appended on the conclusion of each user function call, 
containing information on the calling worker, what type of function was called, 
duration, and the status of any launched applications.

\replace{}{The \libE shutdown process, followed in the case of a standard exit or an exception, aims to extract and save outstanding data where possible, to exit cleanly in different scenarios, prevent hangs, and to maintain the manager's role in coordination, decision making, and logging. The manager captures and handles exceptions that occur both in the manager code itself and exceptions that originate on workers. Worker exceptions are caught and communicated to the manager, which can then report the exception and determine how to handle the error. The default handling is to issue stop signals to all workers, dump the  history and state information to file, and close down the ensemble.}

\subsection{Output and Working Directories}\label{subsec:directories}

\replace{\libE's workers can be directed to call their user functions only after switching into unique
directories on separate file-systems or scratch spaces. This feature is helpful for users who want to
take advantage of high-speed disks, have somewhere to write checkpoint data, or have possibly conflicting 
I/O from their ensemble calculations.}{}

\replace{These output}{\libE's output and working} directories are highly configurable via options specified in the calling script.
\replace{Users should define the location of the \textit{ensemble directory}, which is where all 
workers operate if \libE is initiated on a single node; otherwise, this is the directory name 
on local node storage where each worker operates separately.}{} Workers \replace{}{can} perform each of their computations
in \replace{}{a single} ensemble directory or switch into unique sub-directories. Each output directory can be
set to contain copies or symbolic links of other files or directories; this is helpful for users 
with applications that may modify input files or cases where large initialization files need not be copied for each simulation evaluation.
\replace{}{Such configurability allows users to
take advantage of high-speed disks, configure the location of checkpoint data, and prevent conflicting 
I/O in their ensemble calculations.}

On the conclusion of a \libE routine, workers can copy back their results  to the location that
\libE was originally launched from (usually a user's project directory).

\subsection{Support for User Applications}\label{subsec:supportuserapps}



\libE aims to offer a portable framework for users to provide Python-based functions describing their desired workflow. This includes the capacity
of each worker to launch and manage user applications in a way that is agnostic to
schedulers, MPI runners, launch constraints, and other system variants.

The \libE \textit{executor} provides the user a portable interface to express an
application submission. The executor subpackage is provided
with a base executor class. If the base class is used directly, it will simply
subprocess serial applications locally. The \textit{MPIExecutor} subclass
is most commonly used, which has an extended interface. Most of the following details are based
around this subclass; alternative executor subclasses may easily be added.

The executor can utilize \libE's detection of resources and features 
launch, polling, and termination mechanisms. Launch retries are automatically
used in the case of failures. This interface forms one part of the separation
of concerns between the user's ensemble specification and the mechanism of
launching and managing applications. For example, a completely different launch
system can be swapped out by registering a different executor in the calling script.
One of these, included in the \libE package, is the Balsam executor \textit{BalsamMPIExecutor}
(described in \S\ref{sec:balsam}), which can be used in scenarios where the direct launching  of
MPI applications from the workers is infeasible.


While \libE employs autodetection for
schedulers, node lists, MPI runners, and system factors, there is also a means
for the user to provide this information, overriding detection.

The provision of multiple manager/worker communication substrates, along with
alternative executors, means that most run scenarios can be handled with
minimal changes to the user scripts.



\section{Use Cases and Examples}
\label{sec:uses}
Development work for \libE can take multiple forms. Some researchers are looking to develop efficient 
generators based on state-of-the-art numerical algorithms
with no specific simulator in mind. Others have a simulator (and sometimes a scientific or engineering aim) in mind
and are looking to test or interface with generators.


Here we review four early uses of \libE, each with a particular perspective/objective: 
implementing a generator based on a parallel multistart optimization algorithm (\S\ref{subsec:aposmm}), 
killing a simulation based on its intermediate output (\S\ref{sec:opal}), 
running multi-GPU simulations to achieve a goal within a limited time budget (\S\ref{sec:warpx}),
and
efficiently calibrating a computer model by removing queued calculations before they are executed (\S\ref{sec:cwpcalibration}).

\begin{figure}[!t]
\centering
\includegraphics[trim={0 0 0 21},clip,width=0.35\linewidth]{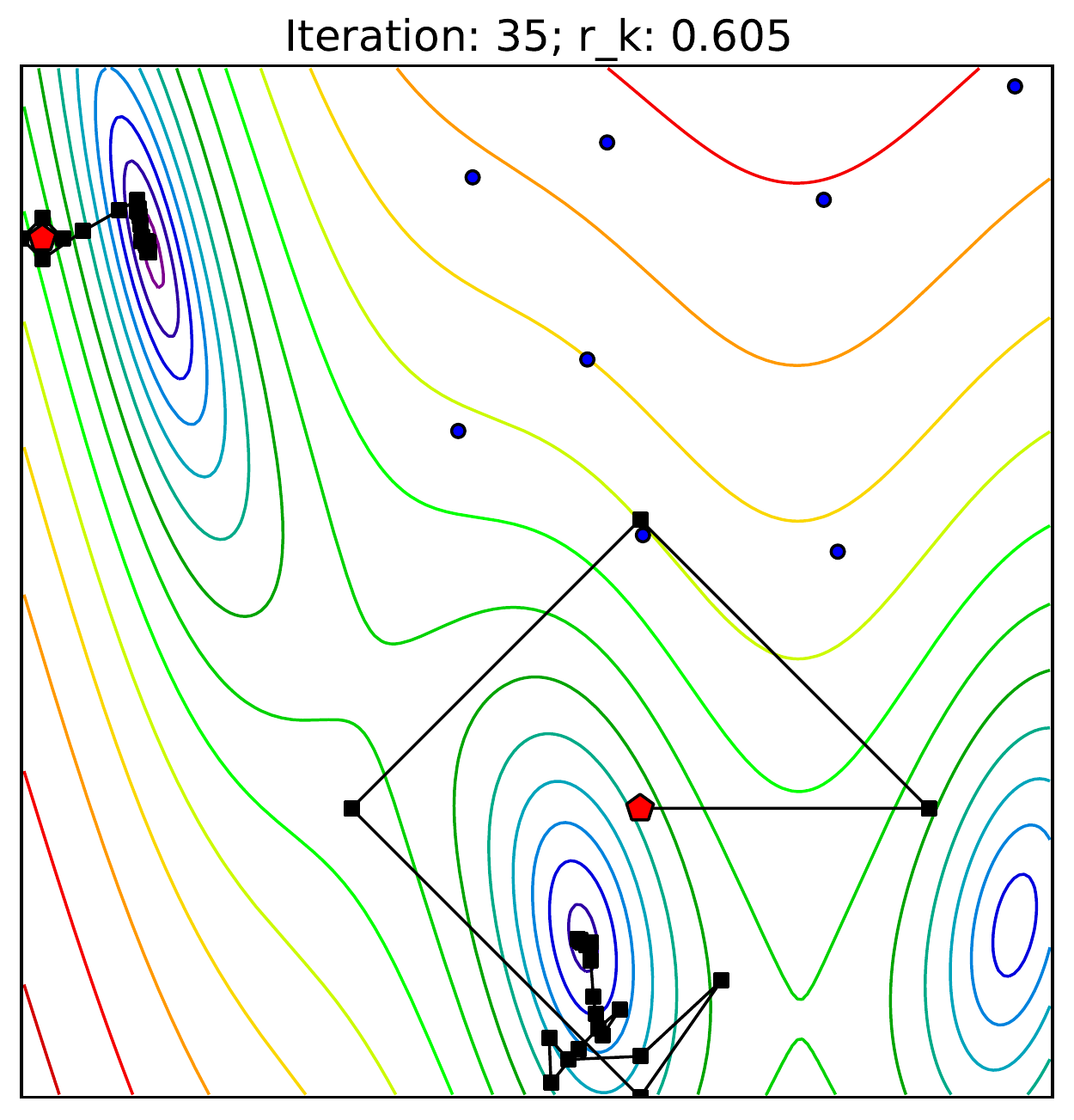}
\caption{Example of the APOSMM generator.
Random samples with
better points in their neighborhood are marked as blue circles; red pentagons
show random points that start runs; and black squares are points arising from
local optimization runs. \label{fig:aposmm}}
\end{figure}

\subsection{Parallel Multistart Simulation-Based Optimization}
\label{subsec:aposmm}
Many computationally expensive simulations of scientific phenomena
have nonlinear, nonconvex behavior with respect to their inputs. In such cases, performing numerical optimization 
can be difficult since the performance of local
optimization methods can depend heavily on 
starting values
and
global optimization methods can require the evaluation of 
undesirably many simulations.

APOSMM~\cite{LW16} is an
optimization solver for finding multiple minima 
using asynchronously parallel objective evaluations.
APOSMM accepts or performs an initial sample of the parameter space and starts
local optimization runs from points that do not have better points in their respective
neighborhood. This neighborhood is adjusted dynamically as outputs from more sets of parameters are
observed. Fig.~\ref{fig:aposmm} shows the contours of a test  objective
function and the points that APOSMM has requested at one snapshot in time. 

In \libE, APOSMM requires the user to declare a local optimization method, with example
interfaces provided for SciPy, NLopt~\cite{nlopt}, and PETSc/TAO
methods, as well as any external routine. That is, the local optimization routine need 
not be in Python, and the \libE regression tests provide examples of
such functionality.


As a generator function, APOSMM allows users to control many aspects of optimization including tolerances
and settings for their local optimization method and how many concurrent local
optimization runs are 
progressing 
at a given time. The asynchronous nature of APOSMM
prevents blocking on a simulation output from 
potentially 
unrelated runs: as soon as a
given run's simulation output is returned from the manager, the next point in
that run can be given. This allows APOSMM to efficiently utilize computational
resources.

\begin{figure}[!t]
\centering
\includegraphics[width=0.5\linewidth]{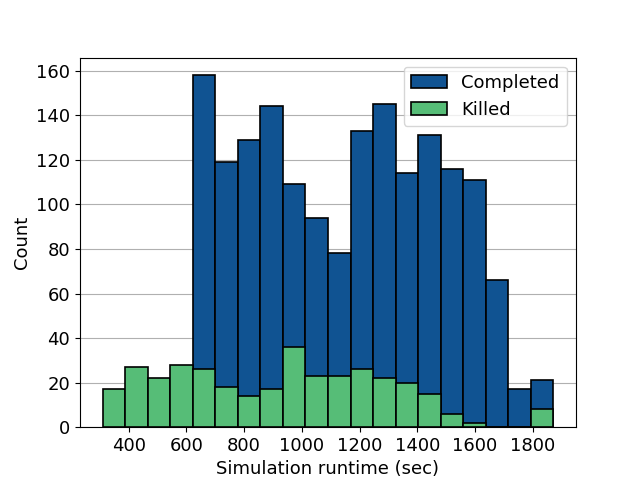}
\caption{Runtime results for OPAL simulations of the LCLS-II beamline
photoinjector in an initial sampling of the 2,046 parameter values. Although this routine used 1,023 workers,
killing runs saved considerable computational resources. The average time for
the 1,696 completed simulations was 1,150 seconds, whereas the average time for
the 350 killed simulations (the completed output of which are not useful) was
941 seconds.
\label{fig:killing_opal}}
\end{figure}

\subsection{Preemption of Photoinjector Sample Configurations}
\label{sec:opal}
    
Researchers at the SLAC National Accelerator Laboratory use OPAL~\cite{opal19}
to simulate the photoinjector for the LCLS-II free-electron laser beam. They
are looking for configurations of laser, magnet, and RF settings that produce a
beam with desired properties~\cite{Lemons20}. 

A typical first step involves sampling over a large domain of allowable
configurations before refining the domain to target specific regions of parameter space. The OPAL simulator reports when particles
are lost from the beamline (an untenable situation in this use case) but does not support termination
as soon as particles are lost. In the \libE simulation function implementation, a feature was added to allow the \libE worker to kill its calculation
based on the contents of the OPAL output file. Figure~\ref{fig:killing_opal} shows the benefit of being able
to kill certain simulations. 
(The killing of simulations is managed by the \libE executor; see \S\ref{subsec:supportuserapps}.)

In this example, the ability to kill certain simulations based on the output
in log files decreased the computational cost required for running
this large collection of simulations. Figure~\ref{fig:killing_opal} shows that,
on average, simulations where particles were lost were killed 3.5 minutes
earlier than the average run times for simulations for which particles were not lost.

\subsection{Designing
Plasma-Based Accelerator Stages}
\label{sec:warpx}
Plasma-based technology is a promising technology to accelerate and focus electron
beams with a smaller device footprint~\cite{Steinke2016}. Designing efficient
plasma-based accelerators includes the tuning of focusing lenses after each
plasma stage. Beamline scientists must determine lens properties, such as
longitudinal position and strength correction factors, in order to optimize beam-quality metrics such as emittance.

High-fidelity simulations of such accelerator designs are critical to
determining realizable designs, but each simulation requires the numerical solution
of both electromagnetic fields and particle movement. WarpX~\cite{Vay2018} is a
particle-in-cell C++ code being developed under the DOE's ECP to address these
challenges on emerging exascale platforms.

Here we illustrate the application of \libE to facilitate multi-GPU WarpX-based
simulations.  As a case study we consider the tuning of two lenses separating
three consecutive plasma stages. A GPU-optimized version of
WarpX~\cite{myers2021porting} was used to simulate each design in a
$192\times192\times7680$ grid with 50,000 particles. This resulted in a typical
simulation runtime of 25 minutes on 2 nodes of the OLCF Summit system (each
node contains 6 NVIDIA Voltas).

\begin{figure}[!t]
\centering
\includegraphics[width=.45\linewidth]{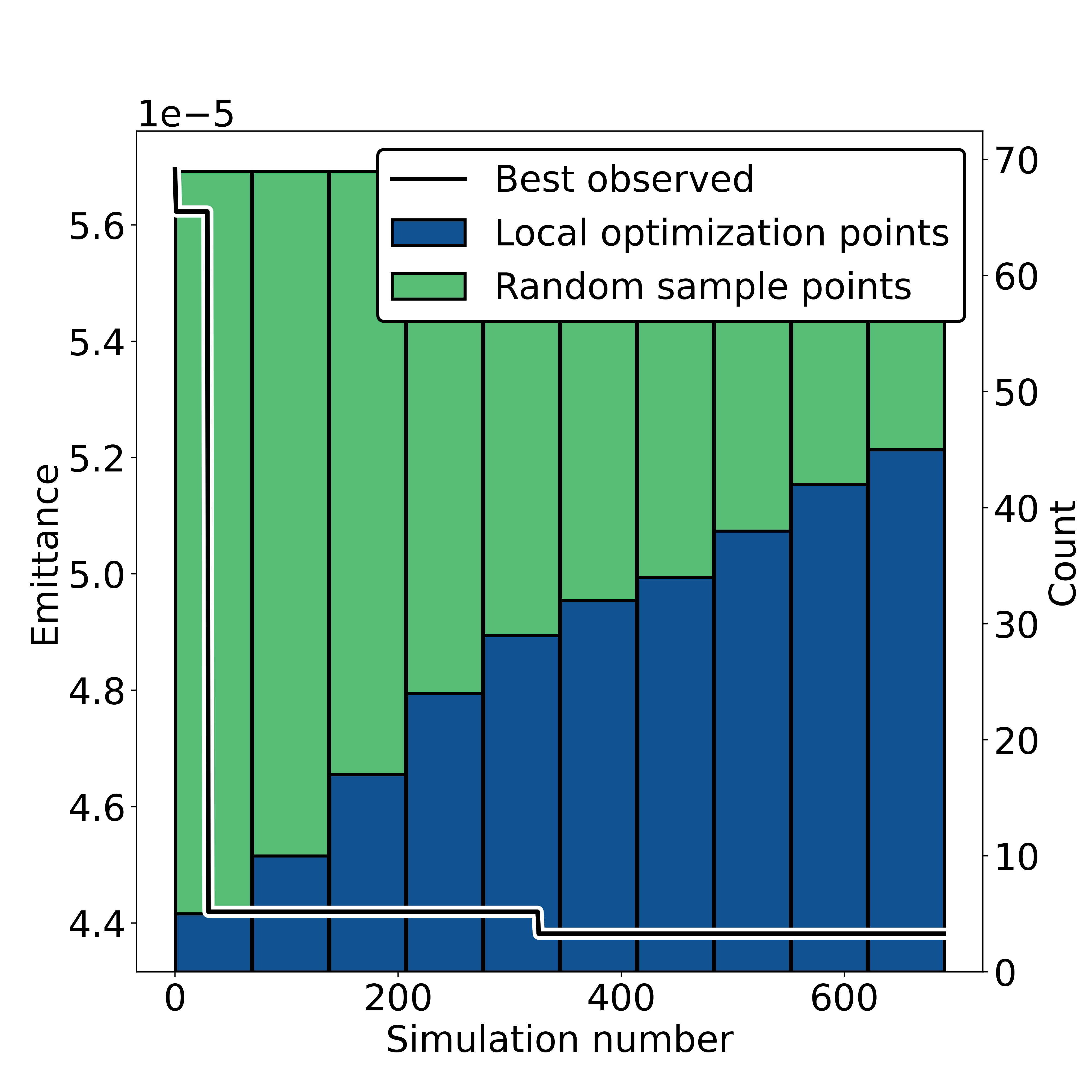}
\caption{Progress of an APOSMM generator in minimizing the emittance from a
beamline. Each simulation is a WarpX run on 12 Volta GPUs. The right vertical
axis represents a visualization (with simulations aggregated into 10 bins) of
the evolving proportion of simulations run for sampling-based exploration
versus those run for local optimization refinement.
\label{fig:warpx}}
\end{figure}

Figure~\ref{fig:warpx} shows the evolution of an emittance-based objective
function value using 47 2-node (12 Voltas) workers. We use an 
APOSMM-based generator that consists of both random sampling (for exploration
of the design space) and local optimization (for refinement within particular
regions of the design space) with the BOBYQA algorithm~\cite{Powell2009a}. We
see that as the optimization progresses, a larger number of the WarpX
evaluations arise from concurrent local optimization runs and hence
refining candidate solutions. By changing generator and allocation functions,
in addition to resources such as the number of workers, \libE can enable
efficient use of connected resources such as architectures where performance is
due primarily  to GPU-based computation.

\subsection{Statistical Calibration of Simulation Models}
\label{sec:cwpcalibration}
A common motif for capability-enhancing  extreme-scale computing is the
calibration of a numerical simulation model to data. Such problems arise throughout
science and engineering and take forms ranging from inverse problems to
supervised machine learning. Statistical calibration seeks model parameters in
cases where the data or model are uncertain, and it tends to require many evaluations of the model. 

One way to solve such problems is to use a Gaussian process
(GP)~\cite{gramacy2020surrogates} emulator to approximate the model output. A
GP
emulator can be constructed by evaluating the model at a well-designed,
predetermined set of parameter values. This is a standard way to construct an
emulator, in part because the process of evaluating a predetermined set of
parameter values scales naturally up to the size of the set. In cases where
such evaluations cannot be performed entirely in parallel, however, determining
parameters to evaluate dynamically can lead to better emulator properties.
Focusing evaluations on portions of a parameter space based on information
obtained from previous evaluations can significantly improve emulator quality
in parameter regions of particular interest for model
calibration~\cite{baker2020analyzing,Binois2018b,BLIZNYUK08}. 

\begin{figure}[!t]
\centering
\includegraphics[width=.4\linewidth]{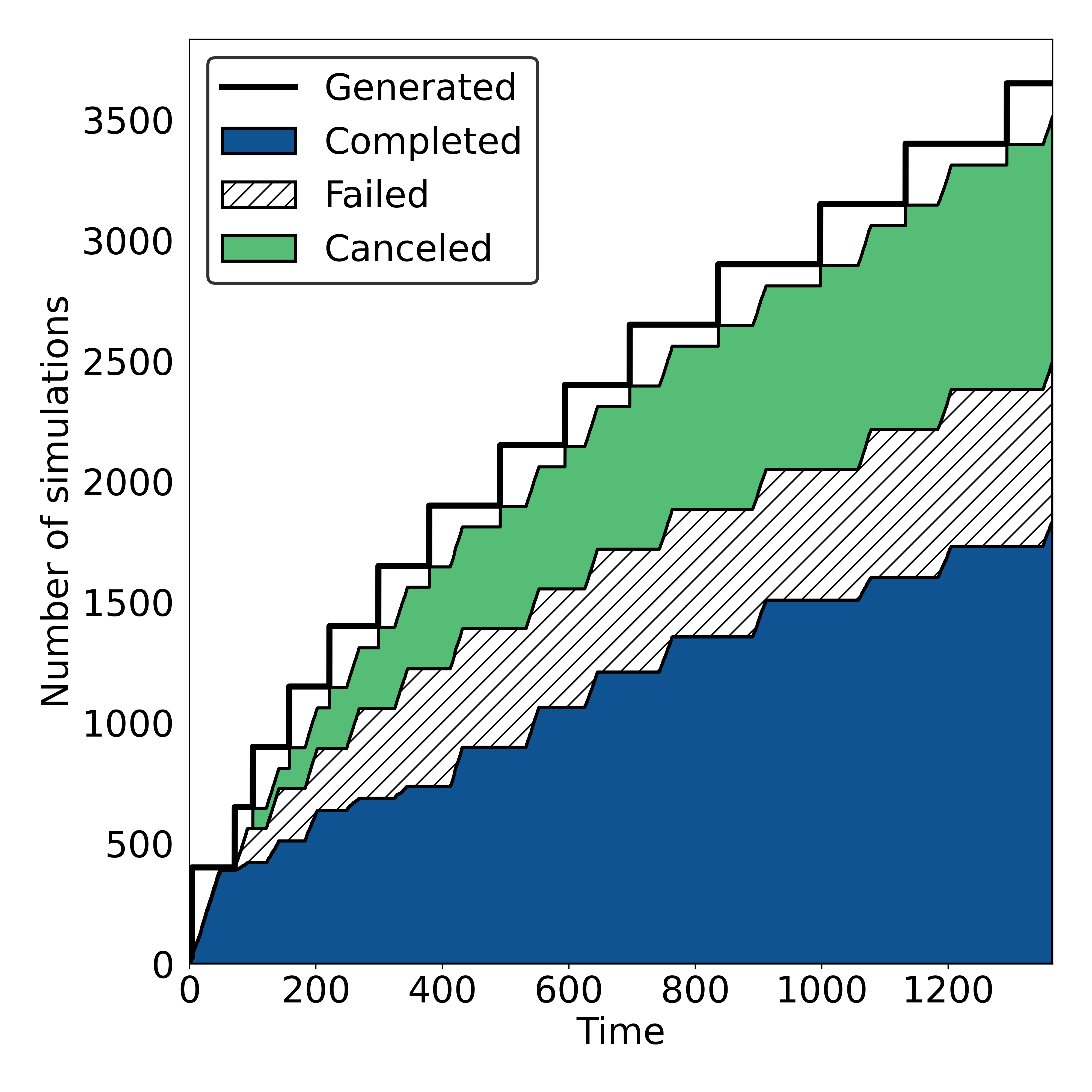}
\caption{Number of borehole simulations that have been generated, completed,
canceled, or marked as failed over the first 14 batches when solving a
statistical calibration problem. Note that this generator function does not mark
points as canceled until 
right before it produces its next set of generated points.
\label{fig:cwpfayans}}
\end{figure}

An active area of research in this area concerns dynamically determining
batches of evaluations to aid the goal of calibration. Unsurprisingly, the
larger the batch size (and thus potentially more parallelizable the
evaluations), the more likely such evaluations are not best suited
for the overall calibration goal.   Determining a batch of evaluations becomes
both more complex and more promising, in terms of overall time to solution,
when considering a capability to preempt/cancel queued evaluations. One \libE
use case involves irregular communication patterns between generator and
manager to preempt specific simulation evaluations. This includes the emulation
capabilities in \texttt{surmise}~\cite{surmise2021,plumlee2019computer}
predicated on generator-directed (e.g., \texttt{surmise-directed})  cancellation of
simulations.


As an illustration of such a use case, we consider a popular statistical test
function involving the calibration of a water flow model based on observed data
in a set of boreholes~\cite{Morris1993}. This test function requires little
time to evaluate, and hence the simulation time can be easily artificially
changed to benchmark a variety of difference computational paradigms.

Figure~\ref{fig:cwpfayans} shows the result of the statistical calibration generator run in persistent mode and creating batches of 
input values
for evaluation. After an initial batch has been processed, the generator
loops, receiving simulation results and, based on particular criteria, decides when to
update the model, create new inputs for evaluation, or request cancellation of previously
issued evaluations that are determined to no longer be useful. 
When evaluations are marked for cancellation, they may be in several states. If
 the evaluation has not yet been issued to a worker, it will not be sent. If
it is already being processed by a worker, a kill signal will be sent by the
manager. The worker can receive this signal and kill a running application
(via the executor). The generator can also handle evaluations that were returned prior to registering for cancellation.


\section{Running on HPC Systems}\label{sec:runningonhpc}

\libE is developed, supported, and tested on systems of highly
varying scales, from laptops to thousands of compute nodes. The aim is to
allow the user, through a minimal set of configuration options, to easily run
an ensemble in the optimal way for the target platform and available resources.

For example, \libE may be run on launch nodes or compute nodes of high-performance computing (HPC)
systems or remotely via TCP; it may use direct or proxy launch of user
applications; and it may schedule resources to workers directly or allow the
system to handle submitted tasks. \libE also simplifies the process
by automatically detecting and interfacing with schedulers (including Slurm,
LSF and Cobalt), MPI runners (including \texttt{mpirun}, \texttt{aprun}, \texttt{srun}, \texttt{jsrun})
and node resources (via environment variables or transparently running a probe task).

\subsection{Configuring the Run}

On systems with a job scheduler, \libE is typically run within a single
job submission. All user functions will run on the nodes within that allocation.
On multinode systems, there are two basic modes of configuring \libE to run
and launch tasks on the available nodes.

\begin{figure}[t]
\centering
\includegraphics[width=0.45\linewidth]{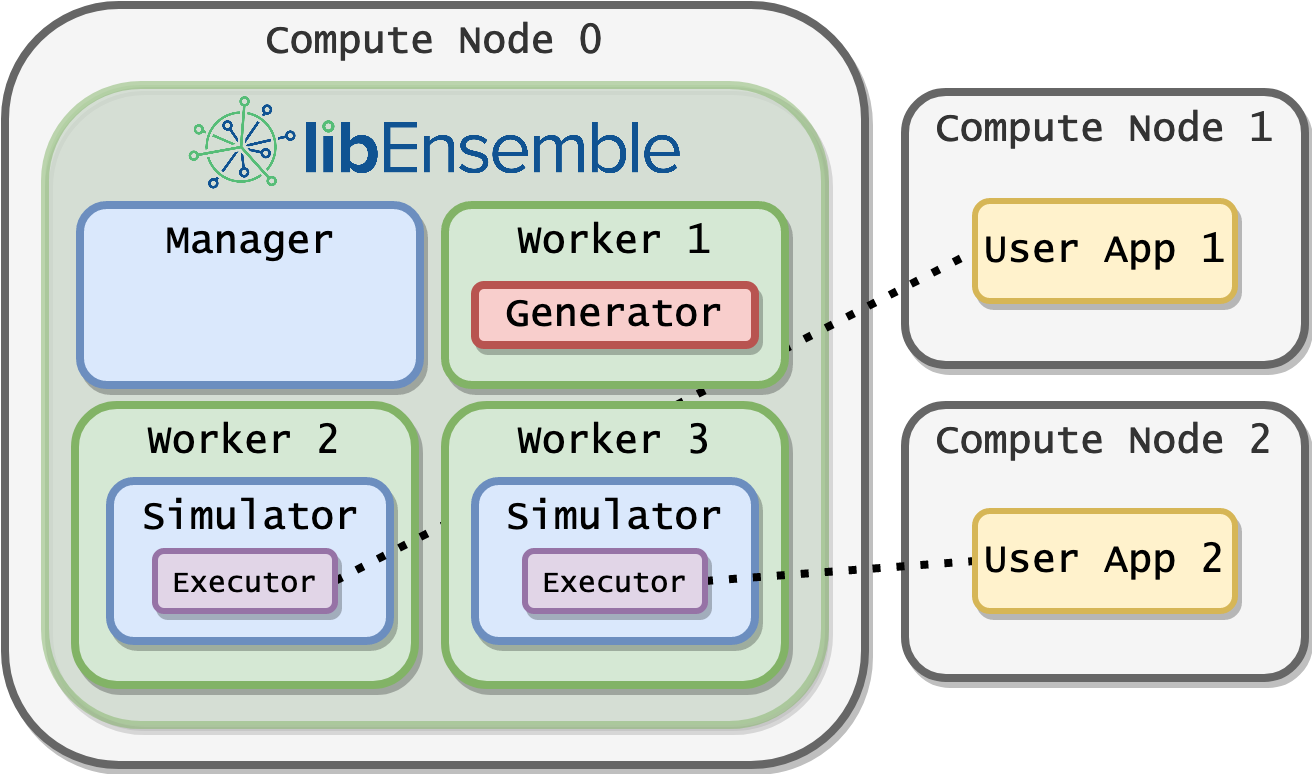}
\caption{Process distribution when launched in central mode.}\label{fig:centralmode}
\end{figure}

\begin{figure}[b]
\centering
\includegraphics[width=.4\linewidth]{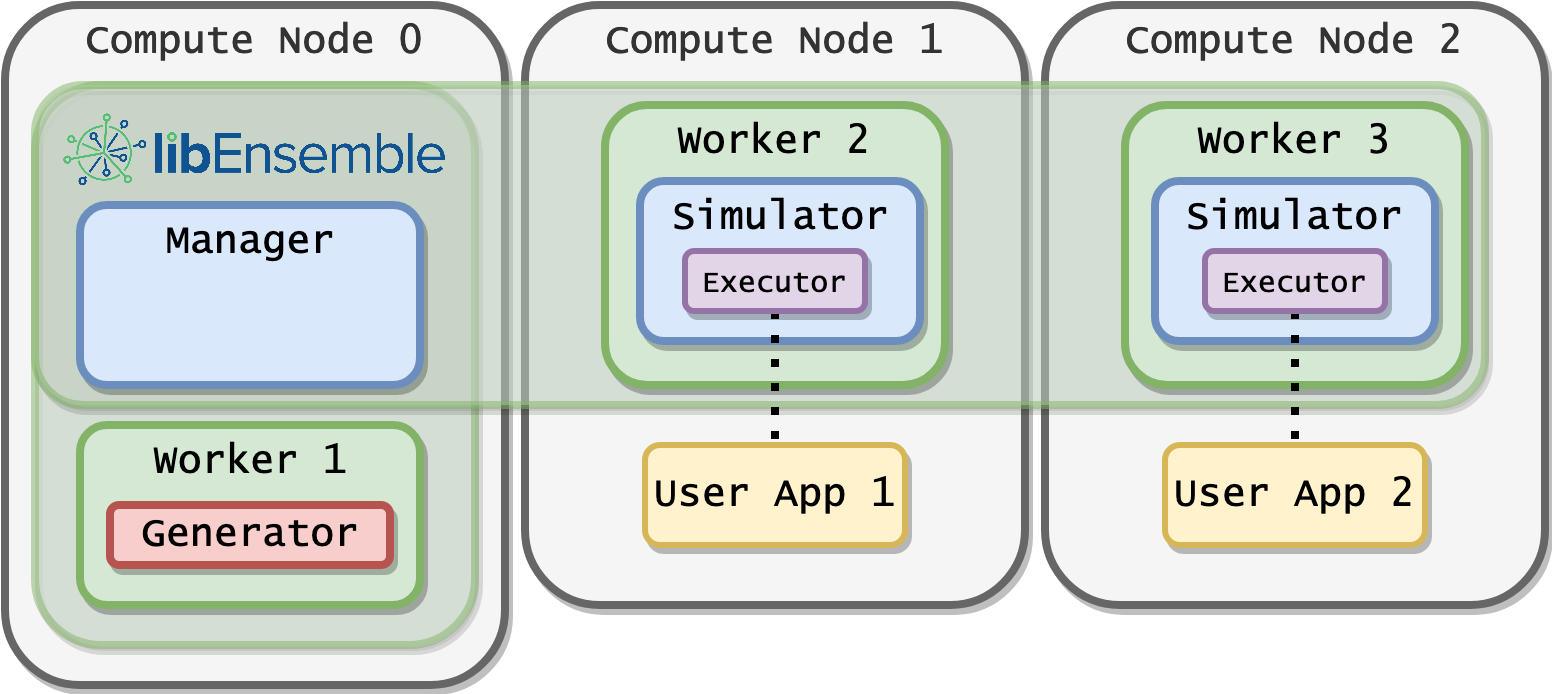}
\caption{Process distribution when launched in distributed mode.}\label{fig:distribmode}
\end{figure}

The first mode we refer to as the \textit{central} mode, where \libE's
manager and worker processes are grouped on one or more dedicated nodes.
Workers launch tasks onto the remaining allocated nodes, as in Fig.~\ref{fig:centralmode}.
Alternatively, in \textit{distributed} mode, the \libE (manager/worker)
processes will share nodes with submitted tasks. This enables \libE,
using the \texttt{mpi4py} communicator, to be run with the workers spread
across nodes so as to be co\replace{}{-}located with their tasks, as in Fig.~\ref{fig:distribmode}.

Configurations with multiple nodes per worker or multiple workers
per node are both common use cases. The distributed approach, for example,
allows the \libE worker to read files produced by the application
on local node storage. HPC systems that  allow only one application to
be launched to a node at any one time prevent distributed configuration.

\libE has built-in resource management that will automatically detect the
nodes available on most systems. Alternatively, the user can provide a  \texttt{node\_list} file in the run directory. While the resource manager
will divide the nodes evenly to each worker by default, more dynamic options are
available (\S\ref{sec:var_resources}). If the
\texttt{central\_mode} option is set, then any node that is running a \libE
manager or worker will be removed from the node-list available for worker-launched
tasks, ensuring that \libE has dedicated nodes for user calculations.

\libE is packaged with example submission scripts, configured to run
either centrally or distributed, for a number of different systems and schedulers.

\subsection{Systems with Dedicated Launch Nodes}

Some large systems such as Summit, Sierra, and Theta have a 3-tier node setup; they have a separate set
of launch nodes (e.g., MOM nodes on Cray Systems), which host user batch jobs
or interactive sessions. Such systems provide a specialized MPI
runner (e.g., \texttt{aprun}, \texttt{jsrun}) that has some application-level scheduling capability.
MPI applications can be submitted only from these nodes.

There are two ways of running \libE on these kinds of systems. The first,
and simplest, is to run \libE on the launch nodes. This is often sufficient
if the workers' simulator and generator functions are not doing too much
work (other than launching applications). This approach is inherently centralized,
with the entire node allocation available for the worker-launched tasks.

To run \libE on the compute nodes of these systems requires an alternative
mechanism (e.g., Balsam, \S\ref{sec:balsam}) for workers to launch tasks. Running \libE on the compute nodes is
potentially
more scalable and will better manage 
simulator and generator
functions 
that contain considerable computational work or I/O.

\subsection{Integration with Balsam}\label{sec:balsam}

Balsam is an HPC workflow management system that comprises a task database (e.g.,
using PostgreSQL), a pilot job launcher that pulls and executes new tasks in real  time,
and an edge service that can submit new batch sessions and transfer data between
systems.

In the typical workflow, Balsam is initiated in the batch submission, where it is
linked to an activated database and adds \libE as the first task. Once
running, \libE can  import the Balsam executor (in place of the regular
executor), and all worker-submitted tasks will be routed to the  Balsam database.
Balsam will take on responsibility for task queuing and scheduling of resources
to each task.

\begin{figure}[b]
\centering
\includegraphics[width=0.4\linewidth]{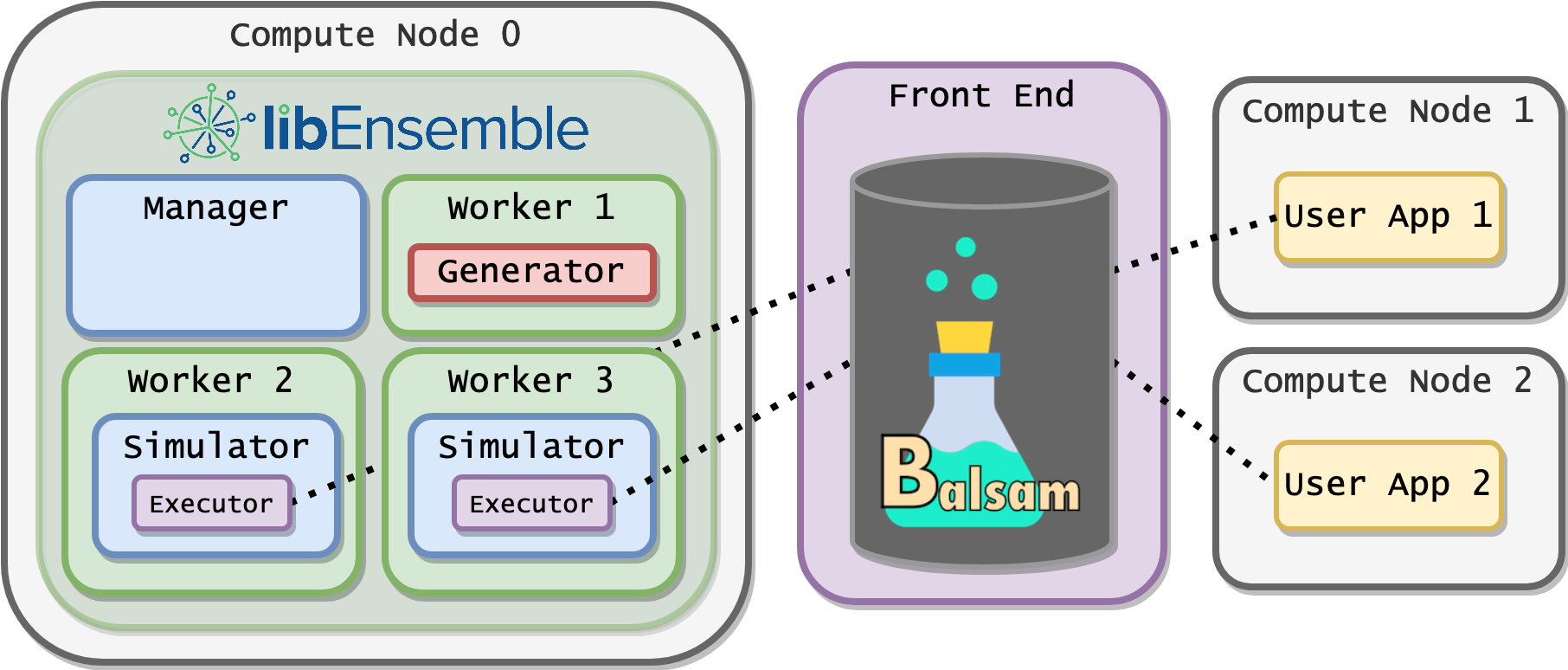}
\caption{\libE can submit tasks to a Balsam instance running on an intermediate
launch node. Balsam dynamically schedules and launches these tasks to the
remaining compute nodes.\label{balsam_fig}}
\end{figure}

Balsam allows the \libE package to be run on compute nodes of systems
such as Theta, thus providing dedicated resources for running any expensive operations
or I/O and ensuring a consistent architecture for any compiled
applications run directly in the user functions. If using the MPI communicator,
\libE's workers can be spread over multiple nodes. Another advantage
is that MPI tasks are decoupled from the parent communicator, so
nested MPI support is not required.

The Balsam edge service component is not necessary for this scenario, but it provides
a capability to automate the orchestration of data transfers between systems and the
allocation of resources/job submission on those systems. This could be employed to
run user tasks in different allocations and across different systems.

\subsection{Resource Management}

\libE has built-in resource management. This entails the detection
of system resources and the allocation of resources to workers.

The necessity for this sort of resource management is system dependent. Some
systems, especially leadership-class supercomputers, support application-level scheduling by default. For example, the \texttt{aprun} command on Theta
will find available nodes within the user's job allocation~\cite{Karo2006}.

Some application schedulers, such as the \texttt{jsrun} system on Summit, support
queuing of applications until a slot becomes available~\cite{JSRUNmulti}.
In this case, the user can easily oversubscribe application runs to the provided
resource allocation.

Furthermore, \libE has the option of submitting via Balsam, which can
also queue and schedule the submitted tasks.

Many systems, however, will simply schedule applications to the same node(s)
unless node information is specified for each application run. Furthermore, if
working at a subnode level (multiple workers submitting tasks to the same
nodes),  node partition information may be required by the worker.

For this reason, and to enable a general, portable approach
to resource management, \libE detects available nodes and cores.
Each worker can then be assigned a node or nodes and, if necessary,
a partition within a node. Automatic resource detection runs by
default but can be disabled or overridden by user-supplied information.

\subsection{Variable-Resource Workers}\label{sec:var_resources}

The resource manager defaults to an even division of resources amongst workers.
However, more flexible and dynamic allocations are possible.

One common case is a generator that runs in place and does not require additional
nodes/cores. For this scenario, especially if the generator runs in persistent
mode,  a worker may be specified as a \textit{zero-resource worker}. The user
can then add an additional worker to make use of those resources.
In Figs.~\ref{fig:centralmode} and \ref{fig:distribmode},
the generator is a \textit{zero-resource worker}.

\begin{figure}[b]
\centering
\includegraphics[width=0.45\linewidth]{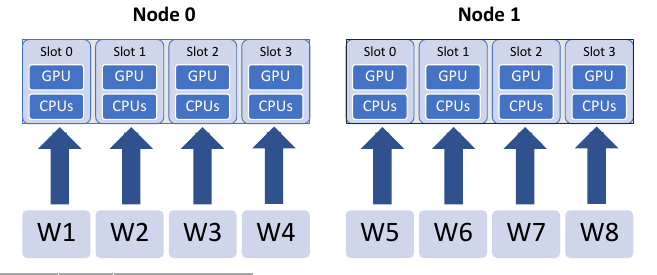}
\caption{Initial fixed mapping of resources to workers.\label{fig:vari_resources1}}
\end{figure}

Ongoing work seeks to update \libE's resource management and
allocation functions to provide better support for the dynamic scheduling of
resources to workers. The current approach takes a fixed initial mapping of workers
to resources, shown in Fig.~\ref{fig:vari_resources1}, where the resources allocated to one worker are defined as a
\textit{resource\ set}. Both manager and workers are aware of this mapping.

\begin{figure}[t]
\centering
\includegraphics[width=0.45\linewidth]{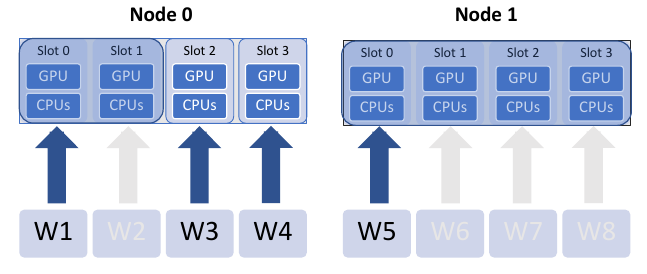}
\caption{After the initial mapping, resources can  be dynamically reassigned to workers.\label{fig:vari_resources2}}
\end{figure}

Resources can then be dynamically reassigned to workers, at the granularity of
\textit{resource\ sets} as shown in Fig.~\ref{fig:vari_resources2}. This feature supports subnode-level workers,
in which case the allocation function's default scheduling policy will find the smallest
partition on a single node. The local partition information is available to each worker,
where it can be accessed directly in user functions or implicitly via the executor. An example
of a simulation function accessing resources information to set the \textit{CUDA\_VISIBLE\_DEVICES} environment
variable is shown in Fig.~\ref{fig:resourcessim}.
The number of resource sets required for each simulation can be determined
in the 
generator
user function. 

\begin{figure}[!b]
\centering
\begin{minted}[
frame=lines,
fontsize=\scriptsize,
]
{python}
from libensemble.resources.resources import Resources
from libensemble.executors.executor import Executor

def my_sim(H, persis_info, sim_specs, libE_info):
    ...
    resources = Resources.resources.worker_resources
    exctr = Executor.executor

    # Convert Python list to comma delimited string
    os.environ["CUDA_VISIBLE_DEVICES"] = \
        ','.join(map(str, resources.slots_on_node))
    num_nodes = resources.local_node_count
    cores_per_node = resources.slot_count  # One CPU per GPU

    # Create application input file
    inpt = ' '.join(map(str, x))

    # Launch application via system MPI runner.
    task = exctr.submit(app_name='six_hump_camel',
                        app_args=inpt,
                        num_nodes=num_nodes,
                        ranks_per_node=cores_per_node,
                        stdout='out.txt',
                        stderr='err.txt')

    task.wait()  # Wait for run to complete

\end{minted}
\caption{Resources and executor accessed by the user simulator function.}\label{fig:resourcessim}
\end{figure}

This approach minimizes communication between the manager and workers. An
alternative approach is for workers to remain unaware of the global mapping. In such a
model, the manager can explicitly assign resources to workers, thus allowing for
greater flexibility: partitions can be arbitrary, and it may be easier to manage
multiple pools of resources, without synchronization. Both of these approaches
are currently being investigated.

Multifidelity ensembles are also currently being explored with collaborators as a
use case where simulations resource requirements vary dramatically.




\subsection{Testing at Scale}

The \libE project conducts an extensive set of small-scale unit and regression
tests using on-line continuous integration (CI) services (Travis CI and GitHub
Actions). However, this is insufficient for testing the interfacing of
\libE with HPC systems.
Therefore, \libE is regularly tested at large scales with a variety of
configurations and test cases on laboratory clusters and leadership computing facilities
located throughout the United States. Scaling tests involve
ensuring successful installation across systems, MPI
initialization, application launches, interoperability with
Balsam, and exception handling at large scales. We typically perform this
testing manually, but recently we have been actively working to automate the
testing process through converting our testing scripts into templates and running
our test suite on high-performance CI instances at Argonne National Laboratory.

\subsection{Support for Exascale Machines}

We anticipate supporting ensemble computations on upcoming exascale leadership-class systems such as Aurora
and Frontier. In the meantime, we are collaborating closely with scalable application (e.g., WarpX) developers who are looking to perform research on such systems.



\section{Discussion and Outlook}
\label{sec:discussion}

The use cases provided here represent a narrow slice of near-term exascale-platform applications. We expect that the aims of \libE will evolve to address diverse applications at even larger, more heterogeneous scales. As an example, we note that controlled environment inverse problems such as those arising from 3D tomographic imaging can already saturate current resources \cite{XCT2020}. Efficiently accounting for imperfect experimental conditions and understanding key uncertainties necessitate
performing not just one expensive inversion but many inversions for ensembles data that are adaptively refined \cite{Kotsi2020,ADLW19}.

\begin{figure}[t]
\centering
\includegraphics[width=0.4\linewidth]{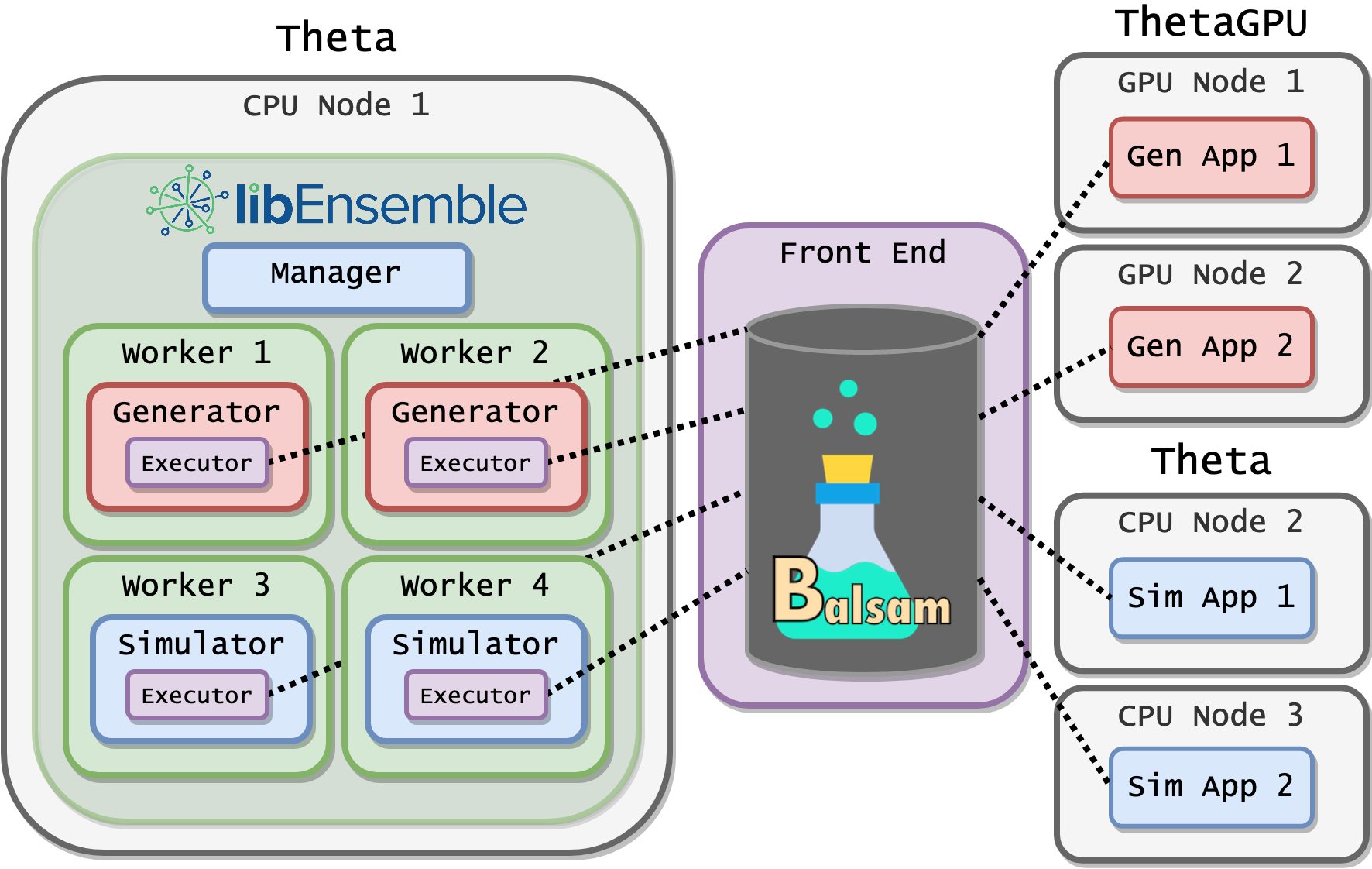}
\caption{Possible workflow using Balsam for cross-system application submissions. In this example, 
generators and simulators submit their respective applications separately to ThetaGPU and Theta.\label{balsam_gpu_fig}}
\end{figure}

With increasing requirements among users to support diverse resources or machines for
ensemble computations, we also anticipate using tools such as Balsam for cross-system application submissions.
Users will theoretically be able to launch \libE on one machine and use the  executor to
launch a subset of applications to a separate machine or set of nodes (each based on a latency that can be statically or dynamically leveraged through an allocation function specification). One such example 
involves a generator launching machine-learning applications to a GPU machine and training them with data 
from simulation applications submitted to a CPU machine, such as in Fig.~\ref{balsam_gpu_fig}.

The ability to use the Balsam-like services to request batch submissions on the fly could also be used for more
efficient scheduling in scenarios with highly uncertain time-scales. For instance, \libE could be run in
a batch session with a long wall-clock time but just enough resources to accommodate the manager and worker processes.
Balsam could request a separate node allocation to accommodate the submitted tasks. Furthermore, new batch
submissions (node allocations) could be created as needed, with queued tasks being maintained in the database
and pulled from any batch jobs created by the Balsam service.

The modular design of \libE means that integration with other pilot systems, runtime environments, and task submission 
mechanisms can be easily achieved by adding an executor.
Also, this modular design has inspired the exploration of the use of a hierarchy of
\libE instances, for example, to mitigate manager communication bottlenecks when
individual simulator, generator, and allocation functions are especially cheap
to evaluate but there is a massive collection of them to be run.

\section*{Acknowledgments}
This research was supported in part by the PETSc/TAO activity within the 
U.S.\ Department of Energy's (DOE's)
Exascale Computing Project
(17-SC-20-SC) and 
by the 
ComPASS and
NUCLEI SciDAC projects within DOE's Office
of Science, Advanced Scientific Computing Research under contract number DE-AC02-06CH11357. 

We gratefully acknowledge computing resources 
operated by the Laboratory Computing Resource Center at Argonne National Laboratory 
and by the DOE
Office of Science Argonne and Oak Ridge Leadership Computing Facilities,  supported under Contracts DE-AC02-06CH11357 and DE-AC05-00OR22725, respectively. 

We are grateful for design help from David Bindel and for inspiration from the PETSc/TAO team. We thank users L\'igia Diana
Amorim, Moses Chan, Remi Lehe, Nicole Neveu, Matt Plumlee, and Maxence
Th\'evenet 
for providing problems and data associated with the presented use cases. 



\bibliographystyle{IEEEtran}
\bibliography{IEEEabrv.bib,references.bib}

\clearpage
\vspace{3em}

\small

\framebox{\parbox{\linewidth}{
The submitted manuscript has been created by UChicago Argonne, LLC, Operator of 
Argonne National Laboratory (``Argonne''). Argonne, a U.S.\ Department of 
Energy Office of Science laboratory, is operated under Contract No.\ 
DE-AC02-06CH11357. 
The U.S.\ Government retains for itself, and others acting on its behalf, a 
paid-up nonexclusive, irrevocable worldwide license in said article to 
reproduce, prepare derivative works, distribute copies to the public, and 
perform publicly and display publicly, by or on behalf of the Government.  The 
Department of Energy will provide public access to these results of federally 
sponsored research in accordance with the DOE Public Access Plan. 
http://energy.gov/downloads/doe-public-access-plan.}}
\end{document}